\documentclass[apj]{emulateapj}

\begin{document}


\title{The Role of White Dwarfs in Cataclysmic Variable Spin-down}


\author{O. Cohen\altaffilmark{1}, J.J. Drake\altaffilmark{1}, V.L. Kashyap\altaffilmark{1}}

\altaffiltext{1}{Harvard-Smithsonian Center for Astrophysics, 60 Garden St. Cambridge, MA 02138}

\begin{abstract}

We study the effect of a white dwarf on the spin-down of a cataclysmic variable system using a three-dimensional magnetohydrodynamic numerical model. The model includes the stellar corona, the stellar wind, and the WD mass and magnetic field. The existence of the WD modifies the system spin-down by physically blocking the stellar wind, restructuring the wind, channeling the wind towards the WD surface, and by modifying the shape and size of the Alfv\'en surface.  The combination of these processes differs among a set of simple test cases, and the resulting angular momentum loss rates vary by factors of 2-3, and by factors of two relative to a test model with a single M dwarf.   While the model employs some simplifications, the results suggest angular momentum loss schemes currently employed in cataclysmic variable studies do not require drastic revision.  Insights are also gained on wind accretion.  We find that efficient accretion switches on quite rapidly with decreasing orbital separation.  Accretion rates depend on magnetic field alignment and should be modulated by magnetic cycles on the M dwarf.  For particular values of white dwarf magnetic field strength, an efficient syphoning of coronal plasma  from the inward facing M dwarf hemisphere occurs.   Wind accretion rates are expected to vary by factors of 10 or more between fairly similar close binaries, depending on magnetic field strengths and orbital separation.

\end{abstract}



\section{INTRODUCTION}
\label{sec:Intro}

The orbital and mass transfer evolution of cataclysmic variables (CVs) and their pre-CV immediate antecedents is though to be largely governed by angular momentum loss through gravitational radiation and to the magnetized wind of the unevolved M dwarf component \citep{Kraft62,Verbunt.Zwaan:81}. 
Angular momentum loss through gravitational radiation is well-understood for the stars in CVs, but there is still no comprehensive theory of spin-down through magnetized winds that is expected to dominate by an order of magnitude or more for periods of $\ga 3$hr, above the CV period gap \citep{Rappaport.etal:83,Andronov.etal:03,Ivanova03}.  

Magnetic braking depends on the wind mass loss rate (MLR) and the large-scale stellar magnetic field \citep[e.g.][]{Weberdavis67,Mestel:68,Kawaler:88}.  Both are extremely difficult to measure for late-type stars and are especially uncertain at the very rapid rotation rates of close binaries where magnetic proxies such as X-ray emission show saturation and supersaturation effects that appear to affect spin-down rates \citep[e.g.][]{Andronov.etal:03,IvanovaTaam03, Barnes03,Wright11}.  Moreover, 
it is generally assumed that magnetic braking  follows the same prescription as for single stars and that the WD makes no difference \citep[e.g.][]{Rappaport.etal:83,Andronov.etal:03,IvanovaTaam03,Davis08}.  While this might be reasonable for WDs with weak magnetic fields that have no direct influence on the magnetospheric structure of the system, it is not  clear that such a situation applies to all systems.  In particular, WDs can have surface equatorial magnetic fields ranging up to 
$10^8$~G that would be expected to have some considerable influence on the magnetically driven wind of a close late-type companion \citep{Schmidt86,Hameury87,Li94a, Li94b,Li95,LiWickramasinghe98}.

Semi-empirical magnetohydrodynamic (MHD) models of the solar wind have now reached a level of realism such that they are able to reproduce quite accurately salient solar wind properties \citep[e.g.][]{Cohen07}.  In previous work, we applied such models to other stars, including rapid rotators and stars harbouring  giant planets
\citep{Cohen09,Cohen10a,Cohen10b}.
Here, we apply the same modelling techniques to begin to tackle the problem of stellar winds and angular momentum loss in close binaries and examine the influence of the companion WD on magnetically-driven stellar winds.  In this pilot study, we consider a synchronized system, such that the WD and late-type star are locked in phase and rotating with the orbital period.


\section{SIMULATION}
\label{sec:Simulation}

This work follows from  \cite{Cohen10c}, who studied the effect 
of a close-in planet on the stellar angular momentum loss to the stellar wind as a function of the 
planetary semi-major axis. They found that the stellar angular momentum loss rate (AMLR) decreases 
once the planetary Alfv\'en surface, which is the ``obstacle'' the magnetized corona and wind feel, starts 
to interact with the stellar Alfv\'en surface.   Here we focus on two other important factors 
that should affect the AMLR of a close binary system:  the mass and magnetic field of a WD, which are much greater than those of the planet. 

We use the solar corona numerical model described by \citet{Cohen07,Cohen08}, which is based on the generic BATS-R-US MHD model \citep{powell99,Toth12}.  The model solves the set of conservation laws for mass, momentum, magnetic induction, and energy with some numerical resistivity, and is driven by surface magnetic field maps that are used both to determine the initial (potential) magnetic field distribution as well as to scale the boundary conditions on the stellar photosphere.  We refer the reader to \cite{Cohen09} and \cite{Cohen10c} for further details 
and the methods we employ to calculate the  model MLR and AMLR.   

For our CV-like system, the donor star is at the center of the simulation box and the WD at $\bar{r}_{wd}=(3R_\star,0,0)$. The numerical MHD code provides a self-consistent 
wind solution for the star, in which the WD is represented by an additional boundary condition for the density, temperature, 
and magnetic field. In addition, the gravitational potential due to the WD mass is included in the MHD equations.   The very small scale of the WD and very high surface magnetic field can cause significant numerical difficulties for the MHD model.   However, the effect of the WD on the system, even out to relatively large radial separations, depends only on its central coordinates, mass and dipole moment, 
and not on the details of its body itself.  We therefore truncate the simulation domain so that the WD boundary (i.e., the WD size) 
is $0.2R_\star$ instead of roughly an Earth size body ($\sim0.01R_\star$). This enables us to have high resolution (small grid size, $\Delta x$) around 
the WD and lowers the magnetic field strength, $B$, near the boundary, which is a key constraint on the simulation time-step.


We adopt a stellar radius and mass of $R_\star=0.5R_\odot$ and $M_\star=0.5M_\odot$, for the donor (the subscript $\odot$ denoting solar values) and 
$M_{WD}=0.75M_\odot$ for the WD.   The star is assumed spherical: Roche lobe geometry and associated distortion of the secondary star is neglected.  
The donor and WD magnetic fields are dipolar with equatorial field strengths of $B_\star=5~G$ and $B_{WD}=777~G$, respectively.  
The latter is equivalent to an equatorial field strength of 
$B\approx 10^6~G$ at the actual WD surface.  Due to numerical limitations,  both the stellar and the WD magnetic fields are an order of magnitude lower than the fields recently observed in M dwarfs \citep{Phan-Bao09,Morin10}, and those used in previous studies of magnetic CV systems \citep{Li94a, Li94b,Li95,LiWickramasinghe98}.  It is not yet clear how 
stellar winds scale with the stellar field strength \citep[e.g.][]{Cohen09,Cohen10b}, but the magnetic balance of our system---the ratio of M dwarf and WD field strengths---approximates a magnetic CV with a WD field $B_{WD}\approx 10^7~G$.

CVs typically comprise  a tidally locked, synchronously rotating secondary, and a non-synchronously rotating WD surrounded by an accretion disk.   We treat a simplified synchronous, diskless case here as a first step; treatment of the more complex asynchronous system requires a much more computationally demanding time-dependent solution beyond the scope of the work in hand.   We omit centrifugal effects and investigate the non-rotating, tidally-locked scenario, neglecting any relative azimuthal motion between the WD and the coronal plasma.   This synchronously rotating WD and secondary is more analogous to the situation in AM~Her-types (despite our adopted WD magnetic field strength, which is an order of magnitude below that required to synchronize the WD rotation; \citealt[e.g.][]{Joss79, KingWhitehurst91, WickramasingheWu91}) and in magnetic pre-CVs, or ``pre-polars'' \citep[e.g.][]{Schwope02,Schmidt05}.  


We performed six case studies, A--F,  probing the effects of WD gravity, field strength and alignment, and orbital separation.
The parameters are summarized in Table~\ref{table:t1}.   We also computed a reference model for the M dwarf alone to judge the influence of the WD on the wind solutions compared to the single star case.  

The simulation box extends from $-45R_\star$ to $+45R_\star$ in each dimension, and a non-uniform grid is constructed so that high-resolution ($\Delta x\approx 0.01R_\star$) is obtained around the two bodies and near the equatorial plane.  
We computed the resulting MLR and AMLR normalized to the values for the reference single star case for which we obtained a total MLR $\dot{M}_0=4.1\cdot10^{12}\;[g\;s^{-1}]$ and an AMLR $\dot{J}_0=7.4\cdot10^{30}\;[g\;cm^{-2}\;s^{-2}]$.


\section{RESULTS \& DISCUSSION}
\label{sec:Results}
\subsection{Model Insights}

In our simulation, the {\it coronal} magnetic field and the stellar wind are computed self-consistently under the disruptive influence of the WD.  The steady-state wind solution for each set of parameters defines the shape and size of the Alfv\'en surface (the collection of points where the wind becomes super-Alfv\'enic). The stellar MLR is simply the integral of the mass flux over this surface.  The AMLR is the integral of the product of the mass flux at the surface (that applies the spin-down torque) with the local radius squared times the rotation rate (see equations 5 and 6 in \cite{Cohen09}).

As we shall see, the WD magnetic field results in significant opening, closing and distortion of the M dwarf field down to the stellar surface in ways that differ according to dipole alignments, WD field and orbital separation.   The MHD wind model employs the 
empirical Wang-Sheeley-Arge (WSA) relation that relates wind speed to the magnetic field expansion rate \citep{WangSheeley90,ArgePizzo00}:   the greater the expansion of the field, the slower the resulting wind.  The different  magnetic field topologies in the cases studied are then naturally expected to affect the resulting wind properties.

\begin{figure*}[h!]
\centering
\includegraphics[width=5.5in]{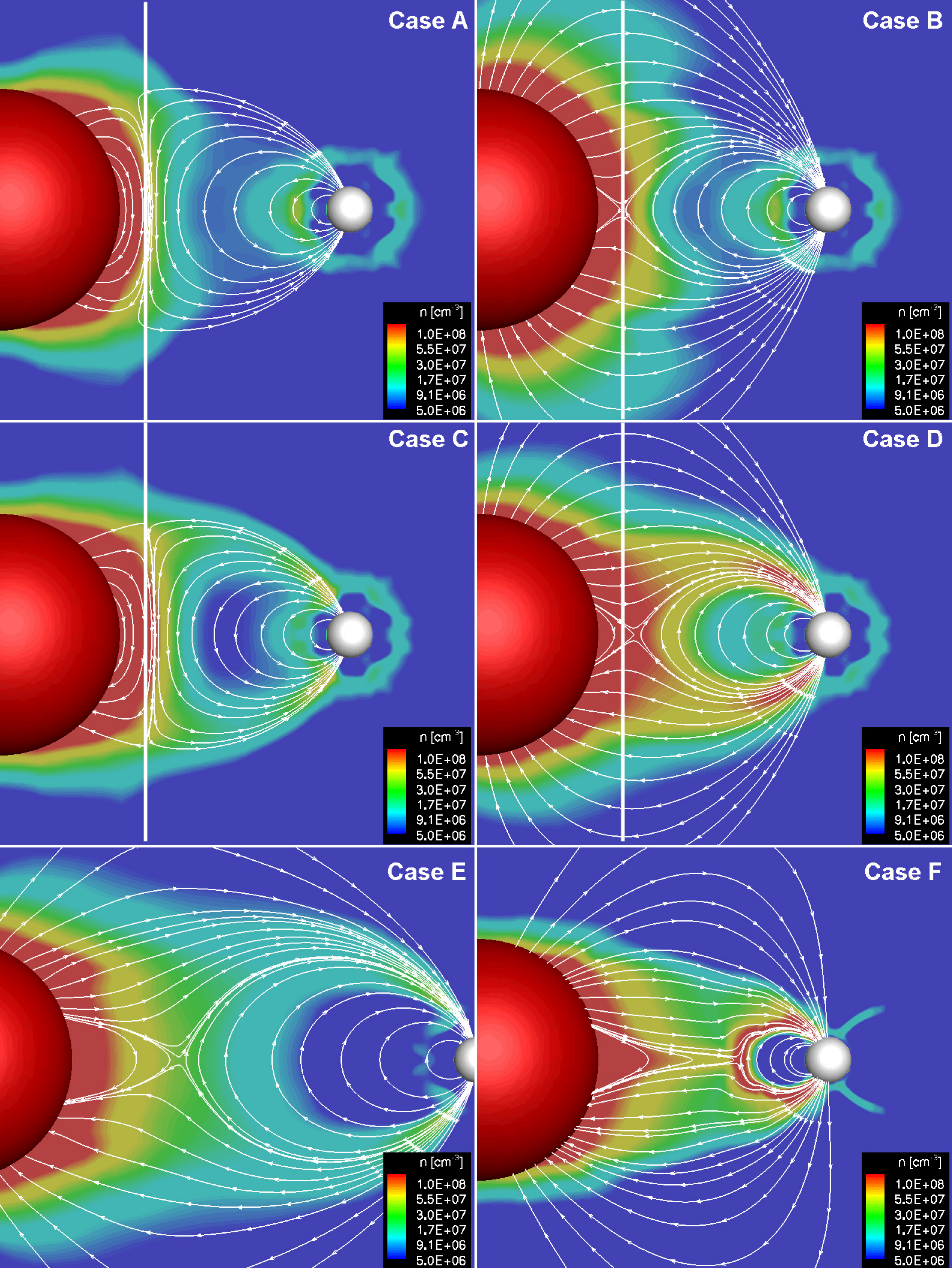}
\caption{The y=0 meridional cut zoomed to the region between the star (red sphere) and the WD (white sphere). Color contours are of number 
density and white streamlines represent magnetic field lines. The solid white line in the two left panels shows the location of the magnetic 
pressure balance of Case A, while it shows the location of the x-point for Case B in the two right panels.}
\label{fig:f1}
\end{figure*}

Figure~\ref{fig:f1} shows the density structure and the magnetic field topology in the region between the donor and the WD for all six cases. There is a region of high density near the star, as well as a ring around the WD, where channelled plasma is trapped within closed field lines. In cases A and C, the WD and donor dipoles are aligned, and a pressure balance point is seen where the two  fields ``collide''.  In the anti-aligned cases (B, D, E, and F), a magnetic (neutral) X-point is seen where the stellar and WD fields reconnect and many field lines connect the two bodies.  This leads to an accretion of stellar wind plasma onto the WD and an expansion of the origin of the high-density wind towards the donor poles. 

The effect of the WD mass is clearly seen when comparing cases A and C, and B with D.  The WD mass focuses the stellar wind to accrete along field lines, increasing the density in the region between the objects.  The pressure balance point (cases A and C) and the magnetic X-point (cases B and D) both move towards the WD when its mass is included.  


\subsection{Mass loss, angular momentum loss and accretion}

Table~\ref{table:t1} summarizes the MLRs, $\dot{M}$, and AMLRs, $\dot{J}$, for the different test cases normalized to the respective single M star values, $\dot{M}_0$ and  $\dot{J}_0$.  The last two columns list the mass accretion rate, $\dot{M}_{acc}$, onto to the WD (normalized to $\dot{M}_0$)
and the total unsigned radial magnetic field at the Alfv\'en surface, $B_o$ (the total magnetic flux opened by the stellar wind).  This flux is not normalized to the reference case since it is dominated by the strong WD field. 

\begin{figure*}[h!]
\centering
\includegraphics[width=6.in]{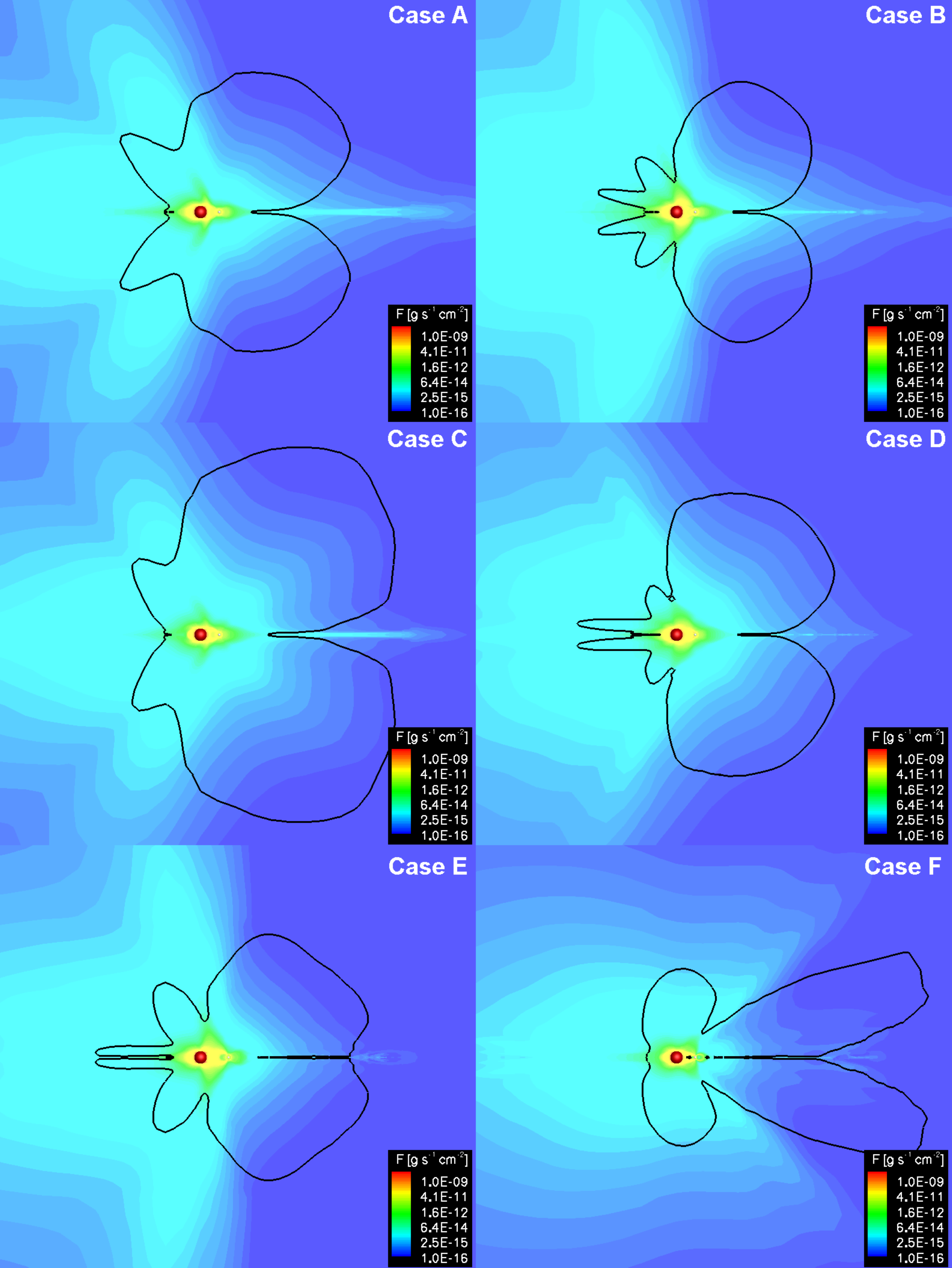}
\caption{The y=0 meridional cut colored with mass flux distribution(in $g\;s^{-1}\;cm^{-2}$). The Alfv\'en surface is shown in solid black line, while the star and the WD are shown as red and white spheres, respectively.}
\label{fig:f2}
\end{figure*}

\begin{figure*}[h!]
\centering
\includegraphics[width=6.in]{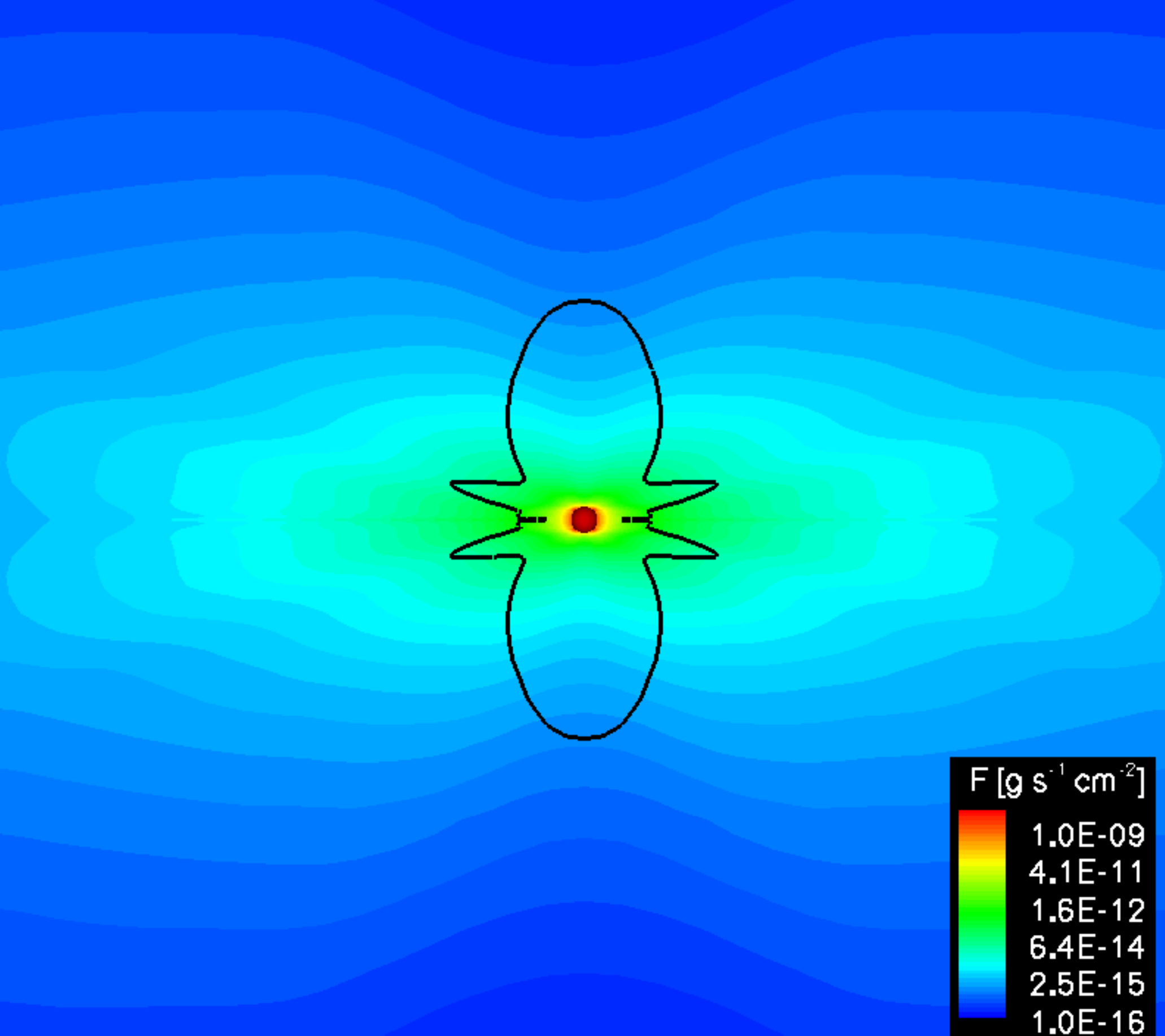}
\caption{The same display as in Figure~\ref{fig:f2} for the reference case without the WD. }
\label{fig:f3}
\end{figure*}

The distribution of the mass flux and the shape of the Alfv\'en surface for the six cases are shown in Figure~\ref{fig:f2}.  The reference case without the WD is similarly displayed in Figure~\ref{fig:f3}.  
In general, the MLR is governed more by the plasma density than its speed.   Wind velocities computed from the model tend not to vary by more than a factor of 2 or so,  as is the case for the solar wind, whereas the wind density is much more inhomogeneous.   The most notable feature in all cases is the reduction of the MLR in the hemisphere occupied by the WD.  

There is a 70--85\%\ reduction in MLR for cases A and B for a massless WD (equivalent to a planet with a very strong field!).  The MLR for aligned dipoles in case A is approximately twice that for the anti-aligned case B.   For anti-aligned dipoles there is a large region of closed field between the stars that prevents wind escape.  Field lines pulled toward the WD increase the field expansion factor at the poles, weakening the wind.  For the aligned case A, the closed field region is much smaller, and field lines are compressed with smaller expansion factors near the poles.  The WD accretion rate for these unphysical massless WD cases is actually negative and is caused by the boundary conditions serving as an evaporating mass source. This boundary effect is removed when the WD mass is turned on.

The reduction in MLR is not as extreme---about 50\%\ for both cases C and D---when the WD mass is included, though for case C there is a large increase in mass accreted by the WD.
When moving the WD further out (case E), the reduction of MLR is only 10\%.  Here the X-point is further from the donor and its closed field corona covers a larger area, shutting off some of the wind.  Coupled with a reduced influence of the WD gravity close to the donor surface, the larger orbital separation leads to a cut in accretion by an order of magnitude.  
In the weaker WD field (case F), the MLR is reduced by 70\% due to the very efficient magnetic coupling that develops between the donor star and WD.  The X-point for this case is much closer to the WD surface than for case~D, leading to smaller field opening angles and efficient "syphoning" of coronal plasma from the donor.

The AMLR is governed by the total MLR through the Alfv\'en surface and by the radial extent of the surface.  The Alfv\'en surface is inflated compared with the single star case, with the inflation being greater for the magnetically aligned fields in cases A and C.  For these cases, the inflation of the surface is more significant than the decrease in MLR, so that the AMLR is 30\% (case A) and 60\% (case C) higher than the reference case.  In cases B and D, the inflation of the surface is less significant than the reduction in MLR, so that the overall AMLR is only about 10\% lower than the reference case. It is interesting to note that by only moving the planet from $3R_\star$ to $4.5R_\star$ in case E this trend is reversed and the AMLR is 25\%\ higher than the reference case.  Despite the significant increase in Alfv\'en surface size, the AMLR in case F is 50\% lower than the reference case.  This demonstrates the importance of the details of the Alfv\'en surface on the AMLR: the extended protrusions in the surface in the left hemisphere where the wind density is reasonably high that are particularly prominent in case D are absent in case E.   Despite the large equatorial extension of the surface in case F in the right hemisphere there is little effect on the AMLR because of the very low wind density in that region.

\section{Conclusions and implications for binary evolution}


We note that there are several ways in which the work presented here could be extended and improved: modeling of an asynchronous rotation of the star and WD; investigation of  field alignments inclined relative to the system spin axis;  inclusion of more complex M dwarf surface magnetic field distributions; treatment of magnetic field strengths stronger by at least an order of magnitude to probe more rigorously the parameter space of magnetic CVs; inclusion of an accretion disk for lower field systems.  

With this caveat in mind, the salient point from the numerical models is that both the wind-driven MLR and AMLR from an M dwarf can be significantly affected by the presence of a WD companion---by factors of 2-3 for the limited parameter range studied.  The alignment of the component magnetic fields can also be important, with differences between aligned and anti-aligned fields amounting to nearly a factor of two in AMLR.  While the differences between the single star and CV-like models are significant, they are not drastic.  We conclude from this pilot study that gross changes to AMLR for spin-down driven by M dwarf winds appear not to be required, especially when noting that different AMLR approaches adopted to date \citep[see, e.g.][]{Knigge11} already harbor uncertainties  as large as, or larger than, the differences between our various models. 
 

The possibility that a strong magnetic field on the WD of a close binary might influence  the magnetically driven wind of the late-type companion was 
suggested by \citet{Schmidt86}, who estimated that the $10^7$~G fields of WD primaries in the magnetic AM~Her-type systems could increase the magnetic lever arm and the AMLR through the secondary wind by factors of 2--5.   
\citet{WickramasingheWu94,Li94a, Li94b,Li95,LiWickramasinghe98} calculated AMLR and other properties of magnetic CVs by superimposing the potential fields of the two components, and using supersonic flow and density profiles which depend on the gravitational potential of the two objects. 
They argued that strong WD magnetic fields could inhibit the donor wind outflow, reducing open field regions and instead capturing the secondary wind and reducing the magnetic braking relative to single stars.  
They predicted that the wind would be completely captured for fields of several $10^7$~G.  



The MHD wind model employed here represents a large improvement over the earlier methods that were based on a potential field approach.   The dependency of the wind properties (velocity and density) on the field structure, as well as the self-consistent evolution of the field structure as it is dragged by the magnetized wind, cannot be treated by potential fields.  The drawback to our models is that 
we are not yet  able to probe the $10^7$--$10^8$~G fields investigated in the earlier studies.   This would be valuable for testing whether the mass loss rates from the M dwarf under such an external field influence are indeed similar to that for a single star.  
However, there are three aspects of our results of relevance for the wind accretion phase of systems with weaker WD fields.  

Firstly, the accretion rate depends quite strongly on magnetic field alignment, as pointed out for stronger fields by \citet{LiWickramasinghe98}, with a factor of 5 increase from case C (aligned dipoles, low accretion rate) to case D (anti-aligned dipoles, higher rate).  During a wind capture phase such as is thought to characterize the ``pre-polars''---magnetic systems not yet filling their Roche lobes \citep[e.g.][]{Schwope02,Schmidt05}---we might then expect to see accretion signatures vary with any magnetic cycles of the donor star.  Such modulation could provide novel insights into magnetic cycles in stars near the fully-convective limit.  

Secondly, we see accretion is reduced by an order of magnitude when the orbital separation is increased from $3R_\star$ to $4.5R_\star$ (case E vs.~case D).  At face value this suggests that wind accretion effectively ``turns on'' as the system evolves to shorter periods much more quickly than would be expected by purely hydrodynamic accretion.  In the  \citet{BondiHoyle44} recipe, the accretion rate varies as the inverse square of the orbital separation \citep[see, e.g.,][]{Matranga11} implying only a factor of 2 difference in rates between cases D and E. 

Thirdly, we find a large increase in mass accretion rate for case F with a {\em weaker} WD field of $10^5$~G.  For this anti-aligned case, effective magnetic coupling between the stars develops, and what would have been closed-field corona is effectively syphoned off onto the WD.  The accretion rate is equivalent to the entire wind of the single star case.  This ``magnetic syphon'' case then mimics the accretion rate of the total wind capture scenario of the pre-polars.   The accretion rate is reminiscent of the $2\times 10^{-13}M_\odot$~yr$^{-1}$ recently found by \citet{Matranga11} for QS ~Vir, a non-magnetic pre-CV that does not appear to fill its Roche lobe.  Such an accretion rate is otherwise difficult to understand under hydrodynamic wind capture, appearing an order of magnitude too high for likely M dwarf wind mass loss rates.

The variation in wind accretion rates by factors of more than 10 within the fairly narrow range of model parameters investigated here suggests that wind accretion rates among pre-CVs could cover a wide range of values.


%


\acknowledgments
OC is suported by NASA-LWSTRT Grant NNG05GM44G. JD and VLK were funded by NASA contract NAS8-39073 to the {\it Chandra X-ray Center}.
Simulation results were obtained using the Space-Weather-Modeling-Framework, developed by the CSEM, at the U. Michigan with funding support from NASA-ESS, NASA-ESTO-CT, NSF-KDI, and DoD-MURI.




\begin{table}[h!]
\caption{Summary of simulated test cases}
\centering
\begin{tabular}{ccccccc}
\hline
Case & $a\;[R_\star]$ & $B_{WD}\;[G] $ & $\dot{M}/\dot{M}_0$ &  $\dot{J}/\dot{J}_0$ & $\dot{M}_{acc}/\dot{M}_0$ & $B_o\;[G]$\\
\hline
A - aligned $B_{wd}$, massless WD  & 3   & $10^6$ & 0.31 & 1.39 & -0.02 & 60.24\\
B - flipped $B_{wd}$, massless WD  & 3   & $10^6$ & 0.14 & 0.90 & -0.03 & 61.15 \\
C - aligned $B_{wd}$, with WD mass & 3   &$10^6$  & 0.51 & 1.61 &  0.09 & 46.43 \\
D - flipped $B_{wd}$, with WD mass & 3   & $10^6$ & 0.48 & 0.92 &  0.55 & 42.09 \\
E - flipped $B_{wd}$, with WD mass & \bf{4.5} & $10^6$ & 0.89 & 1.26 &  0.05 & 53.15\\
F - flipped $B_{wd}$, with WD mass & 3   & $\mathbf{10}^{\mathbf{5}}$ & 0.34 & 0.48 &  0.95 & 23.33 \\
\hline
\end{tabular}
\label{table:t1}
\end{table}


\clearpage

\end{document}